\documentclass[preprint,amsmath,amssymb,aps,showkeys,showpacs]{revtex4}
\usepackage[colorlinks=true, pdfstartview=FitV, linkcolor=blue, citecolor=red, urlcolor=magenta]{hyperref}
\usepackage{graphicx}
\usepackage{latexsym}
\usepackage{amsmath}
\usepackage{amsfonts}
\usepackage{amssymb}
\usepackage{verbatim}
\usepackage{dcolumn}
\usepackage{amssymb}
\usepackage{bm}
\usepackage{float}
\usepackage{subfigure}
\usepackage[english]{babel}
\newcommand{\be}{\begin{equation}}
\newcommand{\ee}{\end{equation}}
\newcommand{\bea}{\begin{eqnarray}}
\newcommand{\eea}{\end{eqnarray}}

\begin{document}

\newcommand{\NITK}{
\affiliation{Department of Physics, National Institute of Technology Karnataka, Surathkal  575 025, India}
}

\newcommand{\IIT}{\affiliation{
Department of Physics, Indian Institute of Technology, Ropar, Rupnagar, Punjab 140 001, India
}}

\title{Null Geodesics and Thermodynamic Phase Transition of Four-Dimensional Gauss-Bonnet AdS Black Hole}

\author{Kartheek Hegde}
\email{hegde.kartheek@gmail.com}
\NITK
\author{Naveena Kumara A.}
\email{naviphysics@gmail.com}
\NITK
\author{Ahmed Rizwan C.L.}
\email{ahmedrizwancl@gmail.com}
\NITK
\author{Md Sabir Ali}
\email{alimd.sabir3@gmail.com}
\IIT
\author{Ajith K.M.}
\email{ajith@nitk.ac.in}
\NITK

\begin{abstract}
Modified gravity theories are of great interest in both observational and theoretical studies. In this article we study the correlation between the null geodesics in the background of a four-dimensional Einstein-Maxwell-Gauss-Bonnet AdS black hole, a modification of Einstein gravity, and it's thermodynamic phase transition. We study the phase structure of the black hole, using the coexistence and spinodal curves, to understand the phase transition in a extended phase space. The imprints of this phase transition features are observed in the behaviour of photon orbit radius and minimum impact parameter with respect to the Hawking temperature and pressure. The change in these two quantities during the phase transition serve as order parameter which characterises the critical behaviour. The correlation shows that thermodynamic phase transition can be studied by observing the effects of strong gravitational field and vice versa.

\end{abstract}


\maketitle

\section{Introduction}
The existence of black hole singularity is one of the most fundamental question in physics. The Penrose cosmic censorship hypothesis asserts that the spacetime singularities need to be hidden from an observer at infinity by an event horizon, which blocks all of the information within it  \citep{Hawking:1969sw, Hawking:1973uf}. Generally all the electrovacuum solutions of classical general relativity are consistent with this conjecture. However, the conjecture does not restrain us from considering black hole spacetimes which are free from singularity, within classical general relativity. In this context, recently proposed theory of four dimensional Gauss-Bonnet gravity is quite interesting one \citep{Glavan:2019inb}. It was demonstrated that for a positive Gauss-Bonnet coupling parameter $\alpha$, the  static and spherically symmetric solution of the theory is free from the much debated singularity problem. The theory is captivating for other reasons too, for example, the obtained black hole solution appears in the setting of the gravity with a conformal anomaly \citep{Cai:2009ua, Cai:2014jea}, and also in the context of quantum corrections \citep{Tomozawa:2011gp, Cognola:2013fva}. However, the black hole solution in four dimensional Gauss-Bonnet theory is attractive, as it is a modified theory of classical gravity, and hence is on an equal footing with general relativity. These captivating features of this novel theory resulted in a surge of various investigations around this theory, including the theoretical aspects, viability of the solution and physical properties \cite{Konoplya:2020bxa, Guo:2020zmf, Casalino:2020kbt, Konoplya:2020qqh, Fernandes:2020rpa,  Lu:2020iav, Konoplya:2020ibi, Ghosh:2020syx, Konoplya:2020juj, Kobayashi:2020wqy, Zhang:2020qam, HosseiniMansoori:2020yfj, Kumar:2020uyz, Wei:2020poh, Churilova:2020aca, Islam:2020xmy, Liu:2020vkh, Konoplya:2020cbv, Jin:2020emq, Ai:2020peo, Heydari-Fard:2020sib, Li:2020tlo, Wei:2020ght, Kumar:2020owy, Hennigar:2020lsl, Mahapatra:2020rds, Shu:2020cjw, Gurses:2020ofy, NaveenaKumara:2020rmi}.

It is well known that black holes are not merely strong gravity systems, but also a thermal systems. Particularly, the establishment of laws of black hole thermodynamics has made the phase transition of these compact objects appealing in every sense \citep{Bekenstein1973, Bardeen:1973gs}. In recent times, anti-de Sitter black hole thermodynamics have gained more interest, as the identification of cosmological constant as the thermodynamic variable pressure, leads to the modification in first law, which has a conventional $PdV$ term \citep{Kastor:2009wy, Dolan:2011xt}. In this extended phase space, AdS black holes exhibit variety of phase transition features, of which, the van der Waals like transition is of great interest \citep{Kubiznak2012, Gunasekaran2012, Kubiznak:2016qmn}. As in the case of a conventional van der Waals fluid, the black hole shows a first order phase transition between two phases, namely, the large black hole phase and the small black hole phase. The authors have studied the thermodynamics of the four dimensional Gauss-Bonnet AdS black hole for both the charged and uncharged cases \citep{Hegde:2020xlv}, and it was observed that a vdW like phase transition exists. Having said so, as the black hole is a gravitational and also a thermal system, it is quite natural to seek a connection between the effects of strong gravity and phase transition. 

It is customary to seek the details of a gravitating object, especially a compact object with strong gravity, by observing the characteristic features of a test particle moving along the geodesics around it. For a particle moving in the vicinity of a black hole, the black hole features are expected to be encoded in the behaviour of the particle motion. These notions are exploited in connecting the unseen attributes of black hole to observational aspects, for example, black hole shadow and quasinormal modes \citep{Cardoso:2008bp, Stefanov:2010xz}. Along with this, the phase transition signature of a charged AdS black hole can be obtained using the quasinormal mode (QNM) studies \citep{Liu:2014gvf}. It was reported that during the van der Waals like phase transition of the black hole, the slope of the quasinormal mode changes drastically, which is an observable phenomenon.

These initial findings motivated to investigate a more concrete relationship between the gravitational and phase transition features of the AdS black holes using the null geodesics \citep{Wei:2017mwc, Wei:2018aqm}. By studying the photon orbits in the background of a charged AdS black hole, the phase transition properties are observed from the behaviour of radius $r_{ps}$ and minimum impact parameter $u_{ps}$ of the circular orbit. The behaviour of $r_{ps}$ and $u_{ps}$ with the Hawking temperature $T$ and pressure $P$, mimics the isobar and isotherms found in thermodynamics counterpart. Below the critical values, the first order phase transition is reflected by these orbital parameters. During the phase transition, these two parameters change by a finite amount, which serves as order parameters to characterise the black hole phase transition, with a critical exponent $1/2$. Originally, this was observed in a charged AdS black holes \citep{Wei:2017mwc}, this correlation between gravity and thermodynamics, via photon orbits, can be seen in different black hole spacetimes, namely, Kerr-AdS \citep{Wei:2018aqm}, Born-Infeld AdS background \citep{Xu:2019yub}, regular AdS black holes \citep{A.:2019mqv}, massive gravity \citep{Chabab:2019kfs}, Born-Infeld-dilaton black hole \citep{Li:2019dai}, five-dimensional Gauss-Bonnet black holes \citep{Han:2018ooi} etc. Related studies in other contexts have also appeared in subsequent works \citep{ Zhang:2019tzi, Bhamidipati:2018yqy,  Wei:2019jve}. In this article we seek a similar correlation for the novel four-dimensional Gauss-Bonnet AdS black hole.

The article is organised as follows. In the next section (\ref{secTD}) we briefly present the 4D Gauss-Bonnet AdS black hole solution and it's thermodynamics. In section \ref{secPT} we investigate the phase transition features of the black hole, wherein, the phase structure is probed using the coexistence and metastable curves. This is followed by section \ref{secgeo}, where we consider the null geodesics on the equatorial plane, and hence obtain the photon orbit radius $r_{ps}$ and minimum impact parameter $u_{ps}$. In section \ref{secphotoncritical}, we study the critical behaviour of $r_{ps}$ and $u_{ps}$, where the order parameters are presented. Finally, we conclude the paper in section \ref{conclusion}.

\section{4D Gauss-Bonnet AdS Black Hole: Metric and Thermodynamics}
\label{secTD}
In this section we briefly present the black hole metric and it's thermodynamics. The $D$-dimensional Einstein-Maxwell-Gauss-Bonnet theory with a negative cosmological constant $\Lambda$ is described by the action \citep{Fernandes:2020rpa},
\begin{equation}
\label{action}
    \mathcal{I}=\frac{1}{16\pi} \int d^Dx\sqrt{-g}\left[ R+2\Lambda +\alpha \mathcal{G} -F^{ab}F_{ab}\right],
\end{equation}
where $g$ is the determinant of the metric $g_{ab}$, $F_{ab}=\partial _a A_b -\partial _b A_a$, is the Maxwell field tensor and $\alpha$ is the Gauss-Bonnet coupling coefficient. The Gauss Bonnet term is given by,
\begin{equation}
    \mathcal{G}=R^2-4R_{ab}R^{ab}+R_{abcd}R^{abcd},
\end{equation}
where $R$ is the Ricci scalar, $R_{ab}$ is the Ricci tensor, $R_{abcd}$ is the Riemann tensor. The cosmological constant is related to the AdS radius $l$ as,
\begin{equation}
    \Lambda = -\frac{(D-1)(D-2)}{2l^2}.
\end{equation}
In four dimensions the Gauss-Bonnet term does not contribute to the dynamics of the system, as the integral over that term is a topological invariant. However, recently a genuine four dimensional Einstein-Gauss-Bonnet gravity was obtained by scaling $\alpha$ as  \citep{Glavan:2019inb},
\begin{equation}
    \alpha \rightarrow \frac{\alpha }{D-4},
\end{equation}
and then taking the limit $D\rightarrow 4$. The spherically symmetric solution for the action (\ref{action}) is,
\begin{equation}
\label{gbsolution}
ds^2=-f(r)dt^2+\frac{1}{f(r)}dr^2+r^2d\Omega ^2_{D-2}.
\end{equation}
In the limit $D\rightarrow 4$ the metric function has the form,
\begin{equation}
\label{metricfun}
f(r)=1+\frac{r^2}{2\alpha} \left(1-\sqrt{1+4 \alpha  \left(-\frac{1}{l^2}+\frac{2 M}{r^3}-\frac{Q^2}{r^4}\right)}\right),
\end{equation}
where $M$ is the ADM mass and $Q$ is the total charge of the black hole. The validity of the theory from which we obtained the above static spherically symmetric solution has been scrutinised in detail in several propositions \citep{Ai:2020peo, Gurses:2020ofy, Shu:2020cjw, Mahapatra:2020rds, Tian:2020nzb, Bonifacio:2020vbk, Arrechea:2020evj}. However, these does not rule out the possibility of having a spherically symmetric solution as it can be obtained from consistent formulations \citep{Lu:2020iav, Kobayashi:2020wqy, Fernandes:2020nbq, Hennigar:2020lsl, Aoki:2020lig}. Therefore we approach the solution (\ref{gbsolution}) as a self-reliant one. Interestingly, this solution also appears in the context of a conformal anomaly gravity  \citep{Cai:2009ua, Cai:2014jea}. 

The horizon of the black hole ($r_+$) is defined by the condition $f(r_+)=0$. Using this condition we obtain the mass of the black hole to be,
\begin{equation}
M=\frac{r_+^3}{2 l^2}+\frac{Q^2}{2 r_+}+\frac{\alpha }{2 r_+}+\frac{r_+}{2}.
\end{equation}
We present the thermodynamics of the black hole in an extended phase space, where the cosmological constant $(\Lambda)$ is treated as the thermodynamic pressure $(P)$, and they are related as $P=-\frac{\Lambda}{8\pi}$. The Hawking temperature of the black hole is associated with the surface gravity $\kappa$, which is,
\begin{equation}
T=\frac{\kappa}{2\pi}=\left. \frac{f'(r)}{4\pi} \right|_{r=r_+}=-\frac{\alpha -8 \pi  P r_+^4+Q^2-r_+^2}{4 \pi  r_+^3+8 \pi  \alpha  r_+}.
\label{Hawking}
\end{equation}
The first law of black hole can be written, considering the GB coupling parameter $\alpha$ to be a thermodynamic variable \citep{Cai:2013qga, Wei:2014hba}, as,
\begin{equation}
\label{firstlaw}
dM=TdS+VdP+\Phi dQ+ \mathcal{A}d\alpha 
\end{equation}
where the potentials $\Phi$ and $\mathcal{A}$ are conjugate to $Q$ and $\alpha$, respectively. Likewise, the thermodynamic volume $V$ is a conjugate to pressure $P$,
\begin{equation}
V=\left( \frac{\partial M}{\partial P}\right) _{S,Q,\alpha}=\frac{4}{3} \pi  r_+^3.
\end{equation}
The entropy of the black hole can be obtained as follows,
\begin{equation}
S=\int _0^{r_+} \frac{1}{T}dM=\frac{A}{4}+2\pi \alpha \ln \left( \frac{A}{A_0} \right),
\end{equation}
where $A=4\pi r_+^2$ is the horizon area and $A_0$ is the integration constant, which has the dimension of $[length]^2$. It is clear that the Gauss Bonnet coupling parameter $\alpha$ modifies the Bekenstein-Hawking entropy-area law.  In general, the black hole entropy is independent of the charge $Q$ and cosmological constant $\Lambda$, therefore, the integration constant can be set as $A_0=4\pi | \alpha |$ \citep{Wei:2020poh}. With this identification, the entropy reads,
\begin{equation}
    S=\pi r_+^2+4\pi \alpha \ln \left( \frac{r_+}{\sqrt{|\alpha|}}\right).
\end{equation}
We emphasise that the black hole entropy has a logarithmic correction, whereas, the thermodynamic volume remains same as the geometric volume. Before concluding the thermodynamics of the black hole, we also mention that, the variables presented above satisfy the Smarr relation in addition to the first law,
\begin{equation}
    M=2TS+\Phi Q-2PV+2\alpha \mathcal{A}.
\end{equation}


\section{Phase Transition of 4D Gauss Bonnet AdS Black Hole}
\label{secPT}
\begin{figure}[t]
\centering
\subfigure[][]{\includegraphics[scale=0.85]{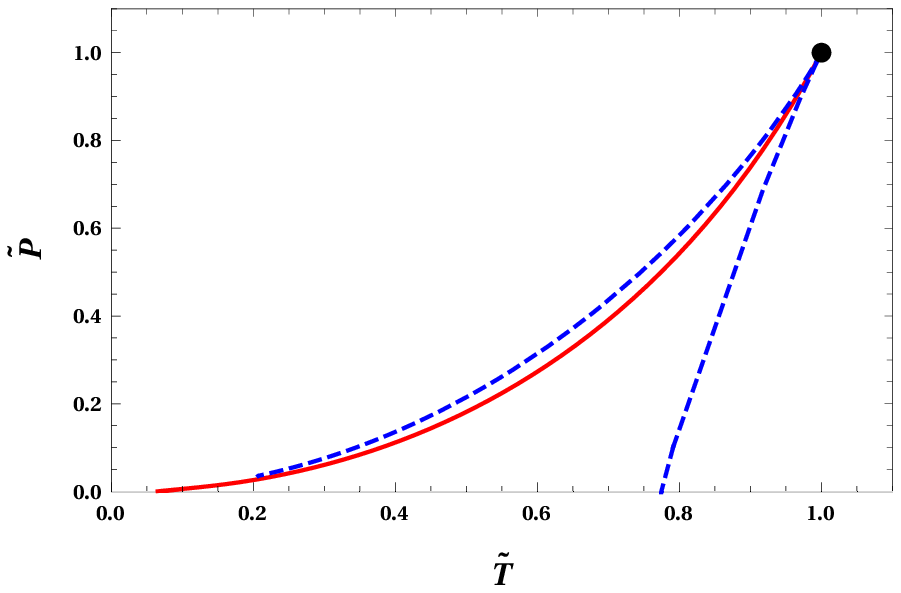}\label{GBPT}}
\qquad
\subfigure[][]{\includegraphics[scale=0.85]{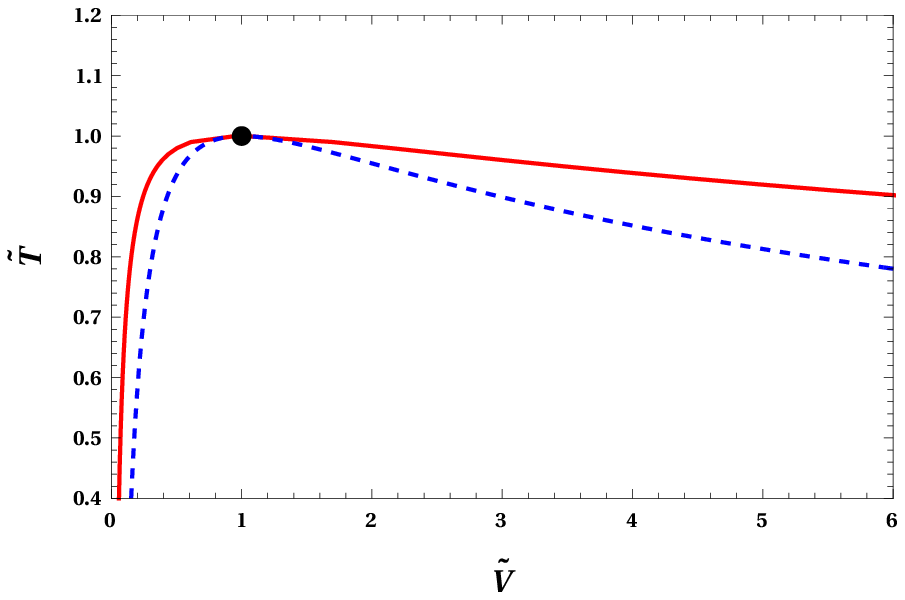}\label{GBTV}}
\caption{The coexistence curve (red solid line) and Spinodal curve (blue dashed line) in $\tilde{P}-\tilde{T}$ and $\tilde{T}-\tilde{V}$ plane. The black dot at $(1,1)$ denotes the critical point. }
\label{GBPTTV}
\end{figure}

The phase transition of the 4D Gauss-Bonnet black hole has been well studied by the authors \citep{Hegde:2020xlv}. Here we recall them to analyse the phase structure using the coexistence and spinodal curves. The state equation of the system is obtained by inverting the expression for Hawking temperature, 
\begin{equation}
P=\frac{Q^2}{8 \pi  r_+^4}+\frac{\alpha }{8 \pi  r_+^4}+\frac{\alpha  T}{r_+^3}-\frac{1}{8 \pi  r_+^2}+\frac{T}{2 r_+}.
\end{equation}
In terms of volume we have,
\begin{equation}
    P=\frac{(6 \pi )^{2/3} (\alpha + Q^2)}{18 \pi ^{1/3} V^{4/3}}+\frac{4 \pi  \alpha  T}{3 V}+\frac{\pi ^{1/3} T}{6V^{1/3}}-\frac{1}{2\times 6^{2/3} \pi ^{1/3} V^{2/3}}.
\end{equation}  
The critical behaviour of the black hole can be easily seen in the $P-V$ isotherms, where a first order phase transition exists between a small black hole phase (SBH) and a large black hole phase  (LBH). This phase transition property is exhibited by both the charged and neutral black holes. The critical point of the phase transitions is determined by using the condition,
\begin{equation}
\left( \frac{\partial P}{\partial V}\right)_{T,Q,\alpha} =\left( \frac{\partial ^2P}{\partial V^2}\right) _{T,Q,\alpha}=0.
\end{equation}
The critical values of the thermodynamic variables are \citep{Hegde:2020xlv},  
\begin{equation}
T_c=\frac{\left(8 \alpha +3 Q^2-\rho \right) \sqrt{6 \alpha +3 Q^2+\rho }}{48 \pi  \alpha ^2};
\end{equation}
\begin{equation}
P_c=\frac{9 \alpha +6 Q^2+\rho }{24 \pi  \left(6 \alpha +3 Q^2+\rho \right)^2};
\end{equation}
\begin{equation}
V_c=\frac{4}{3} \pi  \left(6 \alpha +3 Q^2+\rho \right)^{3/2};
\end{equation}
where $\rho =\sqrt{48 \alpha ^2+9 Q^4+48 \alpha  Q^2}$. Making use of these quantities we define the reduced thermodynamic variables,
\begin{equation}
    \tilde{T}=\frac{T}{T_c} \qquad \tilde{P}=\frac{P}{P_c} \qquad \tilde{V}=\frac{V}{V_c}.
\end{equation}
By observing the phase structure, we can have a better understanding of the phase transition. In the extended phase space, the black hole mass plays the role of enthalpy, which is evident from the first law (\ref{firstlaw}). With this understanding, the Gibbs free energy of the black hole is calculated to be $G=M-TS$, which reads,
\begin{eqnarray}
G=\frac{4}{3} \pi  P r_+^3+\frac{Q^2}{2 r_+}-T \left[\pi  r_+^2+4 \pi  \alpha  \log \left(\frac{r_+}{\sqrt{\alpha }}\right)\right]+\frac{\alpha }{2 r_+}+\frac{r_+}{2}.
\end{eqnarray}
Here $r_+$ is regarded as a function of $(P,T)$ from equation of state. We obtain the coexistence curve in the $\tilde{P}-\tilde{T}$ plane, by using the swallow tail behaviour of the Gibbs free energy. The coexistence expression is also translated into $\tilde{T}-\tilde{V}$ plane. The results are shown in fig. \ref{GBPTTV}. In the $\tilde{P}-\tilde{T}$ plane, the coexistence line (red solid line) partitions the SBH and LBH phases below the critical point. It terminates at the second order phase transition point, above which the phase is supercritical. The figures also display the metastable curves (blue dashed lines) which satisfy, 
\begin{equation}
   ( \partial _V P)_T=0, \qquad (\partial _V T)_P=0.
\end{equation}
The region between the coexistence curve and metastable curve are the metastable phases, namely, superheated SBH and supercooled LBH phases. In the $\tilde{T}-\tilde{V}$ plane the region under the metastable curve corresponds to the coexistence phase of SBH and LBH.

\section{Geodesic equations of motion}
\label{secgeo}
In this section we establish the relationship between the thermodynamics and the null geodesics. Consider a photon which is orbiting the black hole freely on the equatorial plane described by  $\theta =\pi /2$. The Lagrangian which characterises this motion can be directly written from the metric (\ref{gbsolution}),
\begin{equation}
\label{lagrangian}
2 \mathcal{L}=-f(r)\dot{t}^2+\frac{\dot{r}^2}{f(r)}+r^2\dot{\phi}^2.
\end{equation}
Here, the dots represent the differentiation with respect to an affine parameter. The 4D Gauss-Bonnet AdS black hole spacetime has two Killing fields, $\partial _t$ and $\partial _\phi$, which leads to two constants of motion, $E$ and $L$, which are the conserved quantities, the energy and orbital angular momentum of the photon, respectively. The generalised momenta corresponding to the Lagrangian (\ref{lagrangian}) can be obtained by using $p_a =g_{ab} \dot{x}^b$ as,
\begin{eqnarray}
p_t=-f(r)\dot{t}\equiv E\\
p_\phi= r^2\dot{\phi} \equiv L\\
p_r=\dot{r}/f(r).
\end{eqnarray}
The $t$ and $r$ motion of the photon can now be described as,
\begin{equation}
\dot{t}=\frac{E}{f(r)}
\end{equation}
\begin{equation}
\dot{\phi}=\frac{L}{r^2 \sin ^2\theta}.
\end{equation}
The Hamiltonian for the system is obtained from the standard definition, and it vanishes,
\begin{equation}
2\mathcal{H}=-E\dot{t}+L\dot{\phi}+\dot{r}^2/f(r)=0.
\end{equation}
Employing $r$ and $\phi$ motion, we can rewrite the expression for the radial $r$ motion as, 
\begin{equation}
\dot{r}^2+V_{eff}=0,
\end{equation}
with the effective potential given by, 
\begin{equation}
V_{eff}=\frac{L^2}{r^2}f(r)-E^2.
\end{equation}
The photon can only move in the region where $V_{eff}<0$, since $\dot{r}^2>0$. A photon approaching the black hole will be absorbed if it has a smaller angular momentum $L$ and get scattered if the angular momentum is large enough. The absorption and scattering are separated by a critical angular momentum, which defines a unstable circular photon orbit. Expressions governing this orbit are,
\begin{equation}
\label{orbit}
V_{eff}=0\quad , \quad V'_{eff}=0 \quad , \quad V''_{eff}<0,
\end{equation}
where prime denotes a differentiation with respect to $r$. The radial velocity $\dot{r}$ of the photon is zero in this unstable circular orbit. The corresponding value of $r$ is the radius of photon orbit. Expanding the second equation in (\ref{orbit}) we have,
\begin{equation}
2f(r_{ps})-r_{ps}\partial _r f(r_{ps})=0.
\label{aneqn}
\end{equation}
Substituting the metric function $(\ref{metricfun})$ into this equation and solving, we obtain the expression for the radius of photon sphere $r_{ps}$, which is a complicated expression and a function of black hole parameters $(M,Q,P,\alpha)$. Solving the first equation in (\ref{orbit}), $(V_{eff}=0)$, we obtain the minimum impact parameter of the photon as,
\begin{equation}
u_{ps}=\frac{L_c}{E}=\left. \frac{r}{\sqrt{f(r)}} \right| _{r_{ps}}.
\label{upsequation}
\end{equation}
To investigate the correlation between the photon sphere and the black hole phase transition, we observe the behaviour of the radius $r_{ps}$ and minimum impact parameter $u_{ps}$ with respect to the Hawking temperature and pressure, in the reduced parameter space. Apparently this investigation is motivated by the observations in the phenomena of black hole lensing, where the impact parameter $u$ has a close connection with the deflection angle. The deflection angle is  small for a large impact parameter. Yet, in the limit $u\rightarrow u_{ps}$, the deflection angle is unbounded \cite{Bozza:2002zj}.

\begin{figure}[t]
\centering
\subfigure[][]{\includegraphics[scale=0.85]{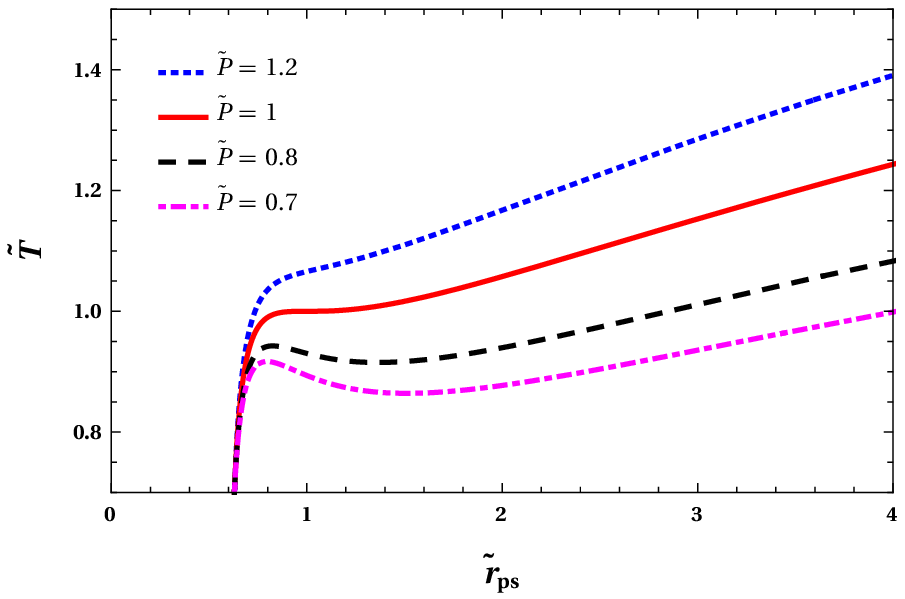}\label{TrGB}}
\qquad
\subfigure[][]{\includegraphics[scale=0.85]{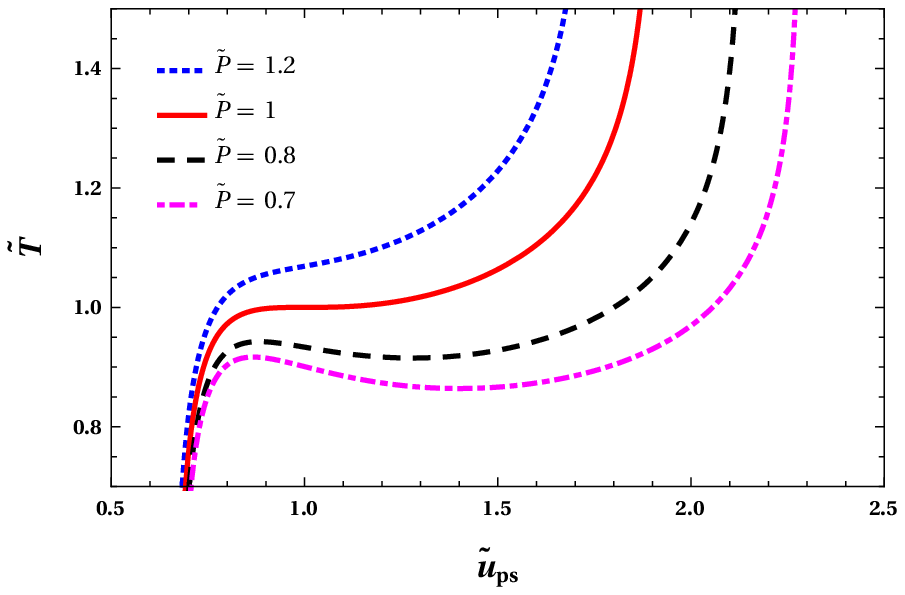}\label{TuGB}}
\caption{The behaviour of photon sphere radius and minimum impact parameter of unstable null geodesic with Hawking temperature in reduced parameter space. We take $Q=1$ and $\alpha=0.5$ }
\label{TruGB}
\end{figure}

\begin{figure}[t]
\centering
\subfigure[][]{\includegraphics[scale=0.85]{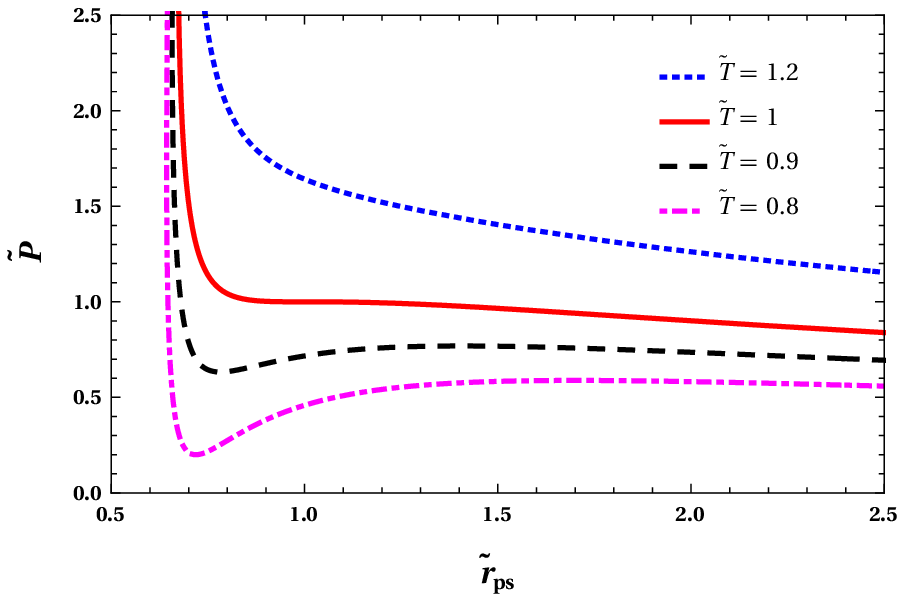}\label{PrGB}}
\qquad
\subfigure[][]{\includegraphics[scale=0.85]{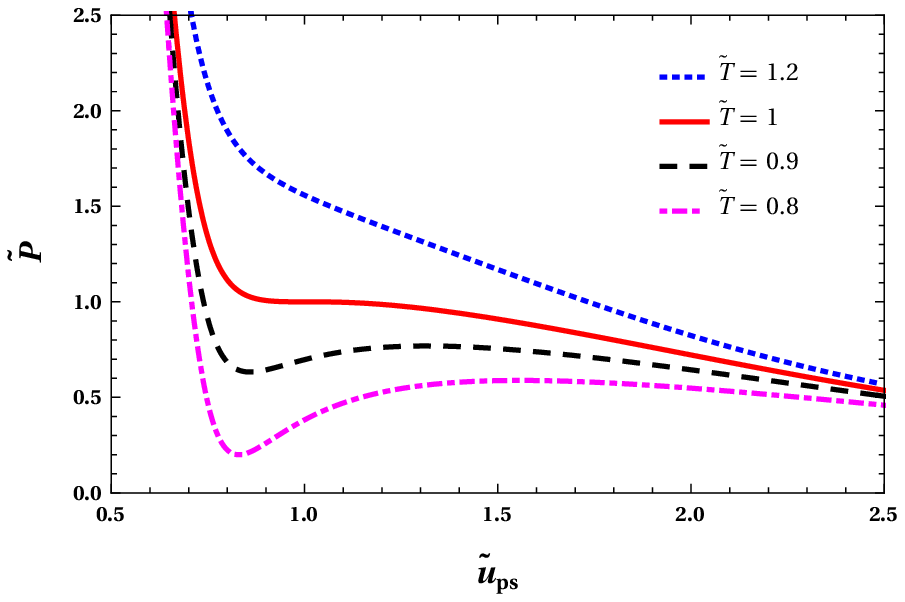}\label{PuGB}}
\caption{The behaviour of photon sphere radius and minimum impact parameter of unstable null geodesic with pressure in reduced parameter space. We take $Q=1$ and $\alpha=0.5$ }
\label{PruGB}
\end{figure}

In fig. \ref{TruGB}, the Hawking temperature $T$ is shown as a function of the photon orbit radius $r_{ps}$ and minimum impact parameter $u_{ps}$, separately, with fixed pressures. The isobars in this figure imply the typical van der Waals like phase transition.  For pressures below the critical value, the isobars first increase, then decrease, and finally increase with respect to the photon sphere radius $r_{ps}$ and minimum impact parameter $u_{ps}$.  In fig. \ref{PruGB} the pressure $P$ is seen as function of $r_{ps}$, first, and then of $u_{ps}$, keeping the temperature as a constant. The behaviour of pressure here (fig. \ref{PruGB}) is opposite to that of temperature (fig. \ref{TruGB}).  For example, when temperature $\tilde{T}$ increases with $r_{ps}$ or $u_{ps}$, the pressure $\tilde{P}$  decreases. In summary, from the behaviour of the photon orbit radius and minimum impact parameter along the isothermal and isobaric curves of 4D Gauss-Bonnet AdS black hole, the van der Waals like phase transition can be clearly identified. This affirms that there exists a correlation between null geodesics and phase transition of the black hole.

\section{Critical behaviour of the photon sphere}
\label{secphotoncritical}

\begin{figure}[t]
\centering
\subfigure[][]{\includegraphics[scale=0.8]{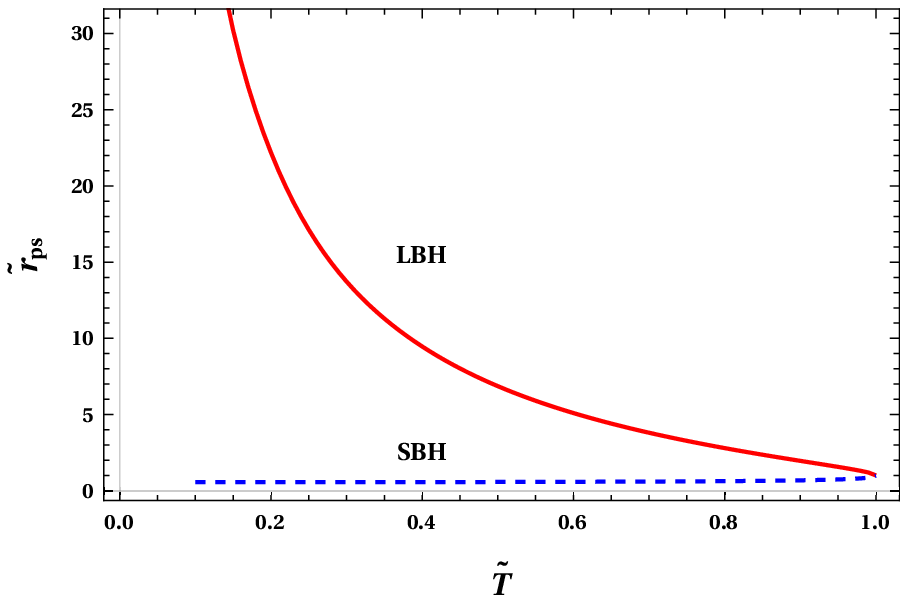}\label{rGB}}
\qquad
\subfigure[][]{\includegraphics[scale=0.8]{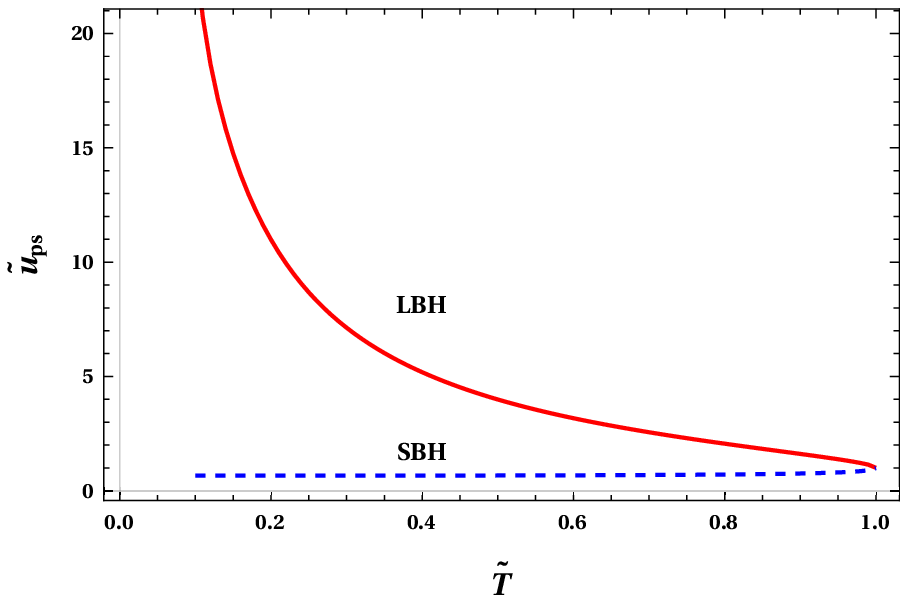}\label{uGB}}
\caption{ The variation of the radius of photon sphere and the minimum impact parameter, for unstable null geodesic, with respect to the Hawking temperature (in reduced parameter space). The SBH (blue dashed line) and LBH (red solid line) meet at the critical point $(\tilde{T}=1)$.}
\label{coex}
\end{figure}

\begin{figure}[t]
\centering
\subfigure[][]{\includegraphics[scale=0.8]{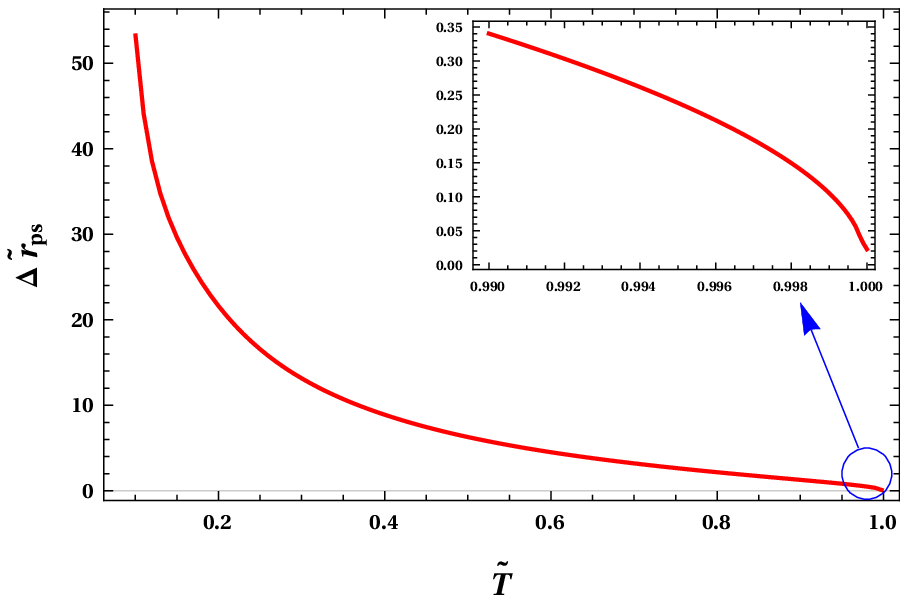}\label{drGB}}
\qquad
\subfigure[][]{\includegraphics[scale=0.8]{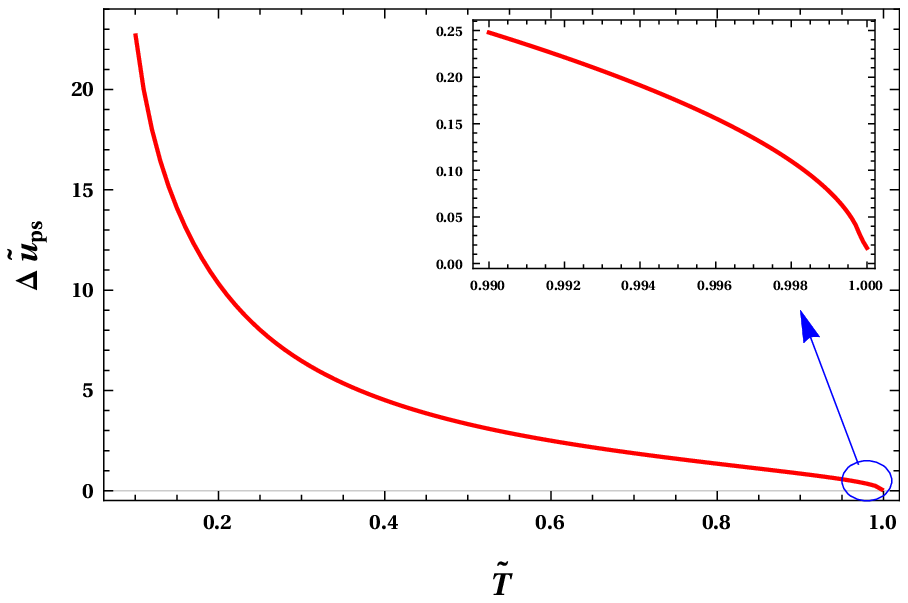}\label{duGB}}
\caption{The change in photon sphere radius and minimum impact parameter of unstable null geodesic during the phase transition of the black hole. The concavity of the curve changes near the critical point, which is shown in an enlarged form in inlets.}
\label{diffcoex}
\end{figure}

The black hole exhibits a first order vdW like phase transition which terminates at the critical point, which corresponds to second order phase transition. As we have seen, there is a connection between the photon sphere and phase transition, it is worth examining the behaviour of change in photon orbit and minimum impact parameter during the phase transition. We construct the equal area law for the $\tilde{T}-\tilde{r}_{ps}$ and $\tilde{T}-\tilde{u}_{ps}$ isobars, similar to the isobars in the $\tilde{T}-\tilde{S}$ plane of the black hole. From the result, we study the behaviour of the photon orbit radius $r_{ps}$ along the coexistence curve (Fig. \ref{coex}). As the temperature increases, the radius $r_{ps}$ for the coexistence LBH phase decreases, whereas for the coexistence SBH phase it increases. The $r_{ps}$ of both coexistence phases attain same value at the critical point $\tilde{T}=1$. Same behaviour is observed for the minimum impact parameter $u_{ps}$. In fig \ref{diffcoex} we display the differences of the quantities $r_{ps}$ and $u_{ps}$ with the phase transition temperature. Both $\Delta r_{ps}$ and $\Delta u_{ps}$  behaves acutely like the order parameter. They have non-zero value corresponding to first-order phase transition and vanish at the second-order phase transition. The behaviour in the neighbourhood of critical point is shown in the inlets. Near the critical point a change in concavity is observed. We numerically obtain the critical exponent of these differences near the critical point to be,

\begin{equation}
\Delta \tilde{r}_{ps} =3.57249(1-\tilde{T})^{0.510839}
\end{equation}
and\begin{equation}
\Delta \tilde{u}_{ps} = 2.54786 (1-\tilde{T})^{0.506096}.
\end{equation}
This behaviour, i.e. $\Delta \tilde{r}_{ps} \sim (1-\tilde{T})^{1/2}$ and $\Delta \tilde{u}_{ps} \sim (1-\tilde{T})^{1/2}$, show that $\Delta \tilde{r}_{ps}$ and $\Delta \tilde{u}_{ps}$ can serve as the order parameters to characterise the black hole phase transition. These results strongly confirm our previous assertion that photon orbits and thermodynamic phase transitions are related to each other.

\section{Concluding Remarks}
\label{conclusion}
In this article we show that the unstable circular photon orbit around the four dimensional Gauss-Bonnet AdS black hole reflects the phase transition information of the black hole. The radius of the photon orbit $r_{ps}$ and the minimum impact parameter $u_{ps}$ are studied in detail. The study establishes a link between the gravity and thermodynamics  in the background of Gauss-Bonnet AdS strong gravity. 

In the first part of the article we presented the thermodynamics and phase transition of the black hole. The phase structure of the black hole is analysed using the coexistence curve and the metastable curves. These curves are the boundaries that separates different stable and metastable phases of the black holes, using which a clear understanding of phase transition features are obtained. The first-order and second order phase transition details are sought in this study, which are influenced by the Gauss-Bonnet coupling parameter $\alpha$. Throughout our study we keep in mind that, the extended phase space thermodynamics features are same for both the charged and neutral Gauss-Bonnet AdS black holes, as it was reported in our previous work \citep{Hegde:2020xlv}.

In the second part of the article, using the Lagrangian of a photon moving freely in the equatorial plane of the black hole we investigated the null geodesics. Using the effective potential, we solve the photon orbit radius $r_{ps}$ and the minimum impact parameter $u_{ps}$ for the 4D Gauss-Bonnet AdS black hole. These two key quantities depend on the black hole parameters, especially the charge $Q$ and Gauss-Bonnet coupling parameter $\alpha$. To establish the relationship between the photon sphere and black hole phase transition we study the behaviour of $r_{ps}$ and $u_{ps}$ along the isobar and the isotherms of the system. The first order phase transition is revealed from these plots. When the pressure or temperature is below the critical value there exists two extreme values for $r_{ps}$ and $u_{ps}$, which coincide to form one extreme point for the critical values of pressure or temperature. Above the critical value of pressure or temperature the $r_{ps}$ and $u_{ps}$ do not exhibit any extremum. Thus they increase monotonically. This behaviour of the photon orbit isobar and isotherm are consistent with that of the black hole thermodynamics. Finally we probe the behaviour of $r_{ps}$ and $u_{ps}$ along the coexistence curve. The two coexistence branches, namely, small black hole and large black hole, have different $r_{ps}$ and $u_{ps}$ values. Their differences $\Delta r_{ps}$ and $\Delta u_{ps}$ serve as order parameters for the black hole phase transition. They vanish near the critical point, which corresponds to the second order phase transition. In the neighbourhood of this critical point, $\Delta r_{ps}$ and $\Delta u_{ps}$ have a critical exponent of $1/2$, which is obtained numerically. Our results show that in the background of Einstein-Maxwell-Gauss-Bonnet AdS spacetime, the black hole thermodynamics can be probed via the strong gravity effects and vice versa.

\acknowledgments
K.H. , N.K.A. and  A.R.C.L. would like to thank U.G.C. Govt. of India for financial assistance under UGC-NET-SRF scheme. 

  \bibliography{BibTex}
  
\end{document}